\documentclass[]{aa}  
\usepackage{orcidlink}

\usepackage{graphicx}
\usepackage{txfonts}
\begin{document}

\title{Solar Orbiter reveals persistent magnetic reconnection in medium-scale filament eruptions}
   \author{
          Song Tan
          \inst{1,2}~\orcidlink{0000-0003-0317-0534}
          \and
          Alexander Warmuth\inst{1}\thanks{Corresponding author; awarmuth@aip.de}
          \and
          Frédéric Schuller\inst{1}
          \and
          Yuandeng Shen\inst{3,4}
          \and 
          Daniel F.\ Ryan\inst{5}~\orcidlink{0000-0001-8661-3825}
          \and
          Daniele Calchetti\inst{6}
          \and
          Johann Hirzberger\inst{6}
          \and
          Takayoshi Oba\inst{6}
          \and
          Artem Ulyanov\inst{6}
          \and
          Gherardo Valori\inst{6}
          }

   \institute{Leibniz-Institut für Astrophysik Potsdam (AIP), An der Sternwarte 16, 14482 Potsdam, Germany
   \and Institut für Physik und Astronomie, Universität Potsdam, Karl-Liebknecht-Straße 24/25, 14476 Potsdam, Germany
   \and State Key Laboratory of Solar Activity and Space Weather, School of Aerospace, Harbin Institute of Technology, Shenzhen 518055, China
   \and Shenzhen Key Laboratory of Numerical Prediction for Space Storm, Harbin Institute of Technology, Shenzhen 518055, China
   \and University College London, Mullard Space Science Laboratory, Holmbury St Mary, Dorking, Surrey, RH5 6NT, UK
   \and Max-Planck-Institut für Sonnensystemforschung, Justus-von-Liebig-Weg 3, 37077 Göttingen, Germany}

   \date{Received/ accepted}

 
  \abstract
    {Solar filament eruptions play a key role in driving space weather, yet their fine-scale evolution remains poorly understood due to observational limitations. Using unprecedented high-resolution observations from Solar Orbiter's Extreme Ultraviolet Imager (105 km/pixel) and Polarimetric and Helioseismic Imager, we reveal persistent magnetic reconnection events in a failed filament eruption. We identify magnetic reconnections between the filament and surrounding magnetic field structures, with both frequency and type far exceeding previous observations. These reconnections significantly affect the filament stability and eruption dynamics, leading to sequential coronal jets and failed eruptions. We propose a 'persistent magnetic cutting' concept, highlighting how persistent small-scale magnetic reconnections cumulatively affect filament stability during its evolution.}


   \maketitle
%

\section{Introduction}
Solar filaments are unique magnetized plasma structures in the solar atmosphere, demonstrating how magnetic fields can confine cold, dense material (temperature $10^{4} \mathrm{K}$, density $10^{11}-10^{12}$ g~$\mathrm{cm}^{-3}$) within the hot, tenuous corona (temperature $10^{6} \mathrm{K}$, density $10^{8}-10^{9} \mathrm{g} ~\mathrm{cm}^{-3}$) \citep{2014LRSP...11....1P,2018LRSP...15....7G,2020RAA....20..166C}. Such magnetically dominated plasma confinement states exhibit rich dynamical behavior, but more importantly, their destabilization can trigger large-scale magnetic energy release, leading to coronal  mass ejections (CMEs) and flares \citep{2011LRSP....8....1C,2012LRSP....9....3W,2015LRSP...12....3W,2020ChSBu..65.3909S,2012ApJ...745L..18A}. These intense events perturb the solar-terrestrial environment, significantly impacting human space activities and ground-based technological infrastructure \citep{2022LRSP...19....2C}.

Decades of observations, theoretical and numerical simulation studies have revealed the physical nature of filament eruptions: it is a process of magnetic disruption and energy release. This process occurs through two main mechanisms: one is magnetic reconnection leading to magnetic topology reconfiguration, including tether-cutting reconnection \citep{2001ApJ...552..833M,2016ApJ...818L..27C,2017ApJ...840L..23X,2018ApJ...869...78C,2021NatAs...5.1126J,2024ApJ...964..125S}, breakout reconnection \citep{1999ApJ...510..485A,2012ApJ...750...12S,2018ApJ...852...98W,2022A&A...665A..51S,2023ApJ...953..148S}, and magnetic flux emergence \citep{2000ApJ...545..524C,2024ApJ...962..149T}; the other involves ideal magnetohydrodynamic (MHD) instabilities, such as kink instability \citep{2003ApJ...595L.135J,2005ApJ...630L..97T} and torus instability \citep{2006PhRvL..96y5002K,2008ApJ...679L.151L,2011RAA....11..594S,2021ApJ...923...45Z,2022MNRAS.516L..12T,2022A&A...665A..37J}. These two mechanisms often work together in different ways, resulting in diverse characteristics such as partial or failed 
filament eruptions \citep{2007SoPh..245..287G}. In particular, magnetic reconnection can not only trigger eruptions, but filaments may also suffer failed eruptions through reconnection with surrounding magnetic fields \citep{2016IAUS..320..221G}. Therefore, accurately identifying and evaluating the role of magnetic reconnection through high-resolution observations is crucial for understanding filament evolution.

\begin{figure*}
\centering
\includegraphics[scale=0.165]{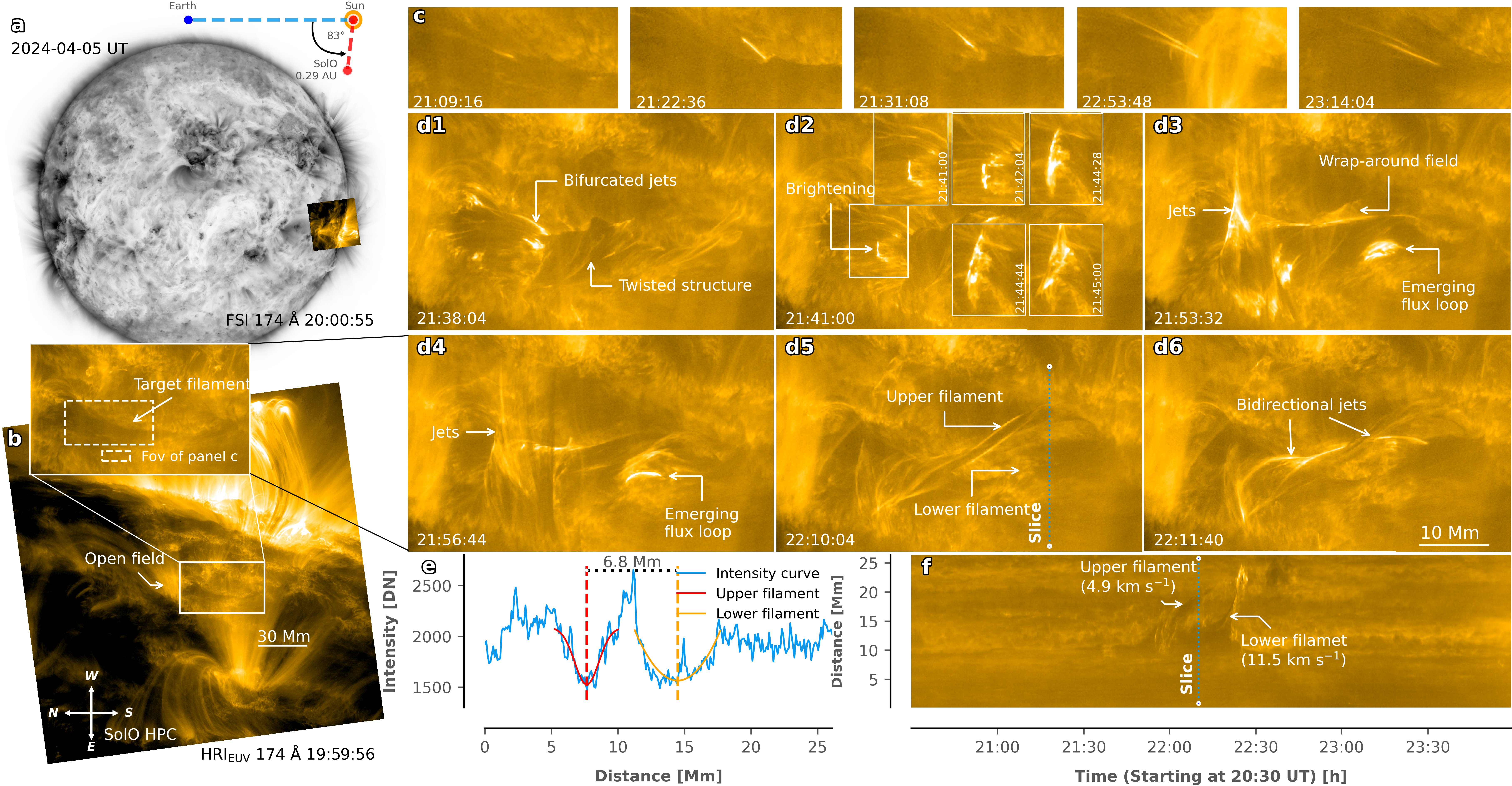}
\caption{\label{fig1}(a): A full-disk image of EUI in 174 \AA\ (inverted greyscale) superposed with $\mathrm{HRI}_{\mathrm{EUV}}$, with the relative positions of SolO. (b): Image of $\mathrm{HRI}_{\mathrm{EUV}}$ (rotated to facilitate the analysis), with the centrally positioned white box (also the FOV of panel (d) and its enlargement labelled with the target filament position). (c): Multiple jets throughout the filament eruption. (d1)-(d6): Filament evolution in $\mathrm{HRI}_{\mathrm{EUV}}$ images. Different features are annotated. The blue dashed line in Panel (d5) indicates the slice position. (e): Intensity profiles (blue) along the slice, where the red and yellow curves fitted with a Gaussian indicate the upper and lower filament position, respectively, and the double-decker filament centers are indicated with dashed lines. (f): Time-space plot of the filament eruption, labelling the upper and lower filaments eruption speed, and the time corresponding to panel (d5).}
\end{figure*}

The European Space Agency's flagship solar mission Solar Orbiter provides breakthrough opportunities to study these phenomena \citep{2020A&A...642A...1M}. Traditional observations from Earth orbit or Lagrange points are limited by single viewpoint and spatial resolution, making it difficult to capture fine structures and rapid evolution processes in the solar atmosphere. By approaching the Sun at an unprecedented distance (0.28 AU), Solar Orbiter's Extreme Ultraviolet Imager (EUI) can observe the solar atmosphere with ultra-high spatial resolution of 100 km/pixel \citep{2020A&A...642A...8R}. This far exceeds the Atmospheric Imaging Assembly (435 km/pixel) on the Solar Dynamics Observatory in Earth orbit \citep{2012SoPh..275...17L}. Although the previous sounding rocket experiment Hi-C achieved observations with 75 km/pixel resolution \citep{2013Natur.493..501C,2014SoPh..289.4393K,2019SoPh..294..174R,2019ApJ...887L...8P,2020ApJ...889..187S,2022ApJ...938..122P}, its observing window of only a few minutes made it difficult to capture complete filament eruption processes.

On April 5, 2024, near perihelion, the High Resolution Imager at 174 \AA\ ($\mathrm{HRI}_{\mathrm{EUV}}$) of EUI observed continuously for 4 hours (19:59-23:59 UT, 16-second cadence). Meanwhile, the High Resolution Telescope (HRT) of the Polarimetric and Helioseismic Imager (PHI) also obtained 8 frames of magnetic field data (20:00-23:30 UT, 30-minute cadence) \citep{2020A&A...642A..11S}, fully covering the filament eruption process. 
This unique dataset allows us, for the first time, to study the  magnetic reconnection processes in filament evolution with unprecedented resolution.

\section{Results}
\subsection{Multiple jets throughout the eruption}

Solar Orbiter provided a unique vantage point at 0.29 AU from the Sun, 83° from the Earth-Sun line on 5 April 2024 (Fig. \ref{fig1}(a)). $\mathrm{HRI}_{\mathrm{EUV}}$'s FOV was at the western solar limb, so the target filament (W150S18, about 40 Mm in length) was completely hidden from Earth's view. Fortunately, $\mathrm{HRI}_{\mathrm{EUV}}$'s 4-hour observing window completely covered the filament eruption (21:00-23:00 UT). Fig. \ref{fig1}(b) shows the $\mathrm{HRI}_{\mathrm{EUV}}$'s FOV and marks the target filament. In $\mathrm{HRI}_{\mathrm{EUV}}$ images, the filament was located between several active regions with extended nearby open field  to its left.

Precursor phenomena, such as magnetic cancellation and corresponding brightening, often precede filament eruptions. Multiple jets disturbed the filament (Fig. \ref{fig1} (c), not all jets could be shown; see movie starting at 21:39:08 for other examples), which we consider to be the precursor of this filament eruption. Following one particularly bifurcated jet (Fig. \ref{fig1} (d1)), the filament exhibited a clearly twisted structure, revealed by hot plasma flows illuminating its fine structure \citep{2013ApJ...770L..25L,2014ApJ...784L..36Y,2019ApJ...883..104S,2022MNRAS.516L..12T,2023MNRAS.520.3080T}. With the filament activated, we observed a bright loop structure below the filament (Fig. \ref{fig1} (d3) and (d4)). The bright loop acts as the signal of emerging flux loop, and its location is consistent with the location of multiple jets initiation, suggesting that the reconnection of the emerging flux loop with the filament may be the cause of these multiple jets. Based on the HRT LOS magnetogram at 20:30 UT, we also detected positive polarity emergence at the same location as the emerging flux loop (Appendix \ref{A2}). The magnetic flux emergence rate was about $2.62\times 10^{18}$ Mx $\mathrm{hr}^{-1}$, which is slightly lower than previously reported emerging flux loop events \citep{2018ApJ...861..108Z,2022MNRAS.516L..12T}. Given  temporal resolution (30-minutes) and the limb projection in this filament, this rate falls within a reasonable range.
At the same time (Fig. \ref{fig1}(d3)), we also identified several field lines wrapped around the filament that will play an essential  role in the subsequent filament eruption.

\begin{figure}
\centering
\includegraphics[scale=0.87]{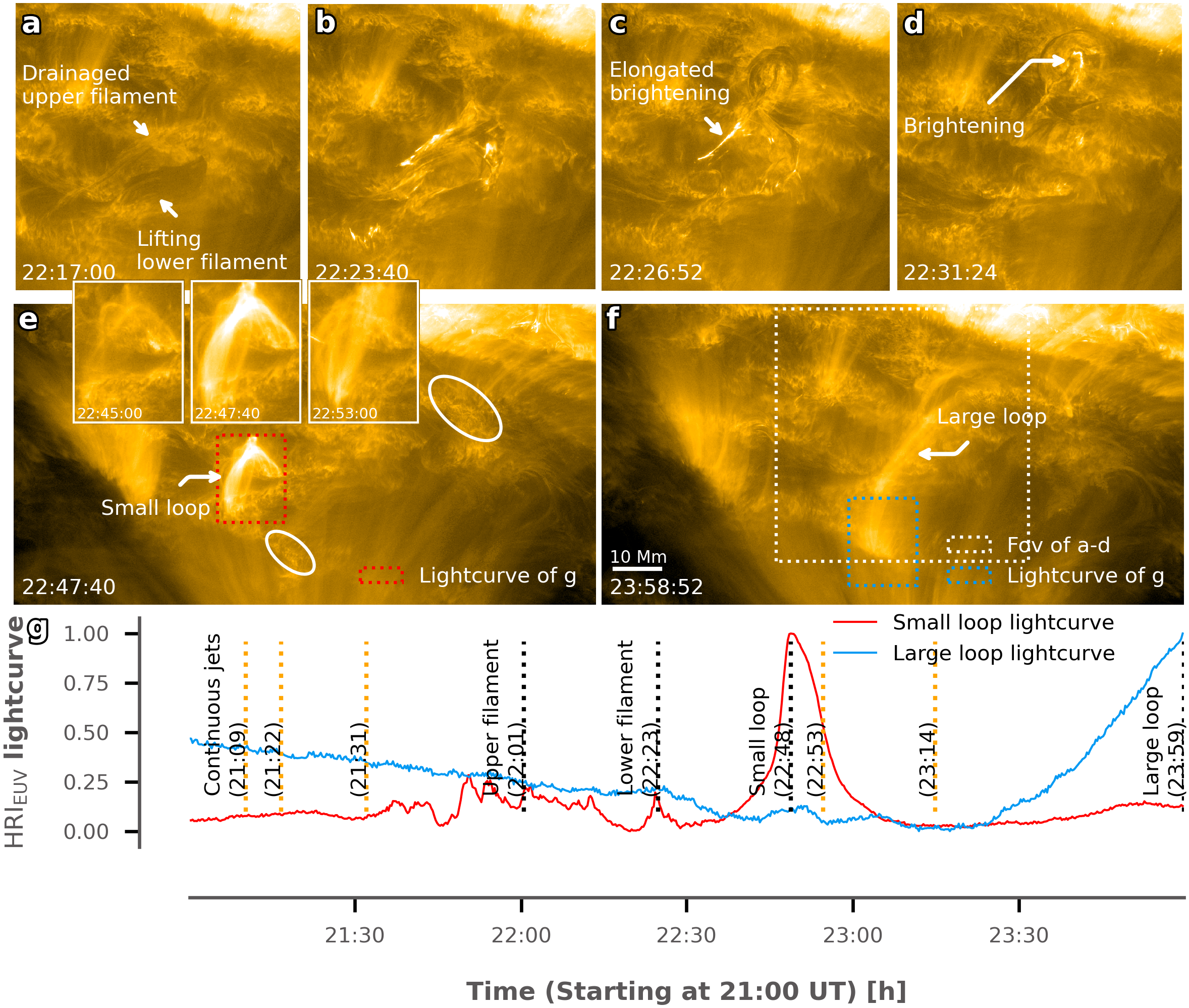}
\caption{\label{fig2}(a)-(d): The LF eruption process, with FOV marked in white dashed boxes in panel (f). (e): The small post-flare loop produced by the LF eruption, with the loop region represented by the red dashed box. (f): Large loop  visible after the LF eruption, with the left footpoint region represented by blue dashed box. (g): Normalised lightcurves corresponding to the red and blue dashed boxes with the moments of the small and large loops marked, as well as moments of multiple jets and filament eruptions.}
\end{figure}

\subsection{Reconnection of the filament and open field}
Prior to the main eruption, a brightening appeared close to the filament's left footpoint near the nearby open field (Fig. \ref{fig1}(d2)). The brightening then began to extend to the sides (up and down) into a bidirectional jet until it evolved into a collimated coronal jet (Fig. \ref{fig1}(d2)). The observed jet characteristics initially appeared consistent with mini-filament-driven coronal jets \citep{2011ApJ...735L..43S,2015Natur.523..437S,2015ApJ...806...11M,2020ApJ...896L..18S,2023FrASS..1017870S,2024ApJ...969...48P,2021RSPSA.47700217S}, where the erupting filament and the open field are reconnected, and the filament material  escapes into interplanetary space along the newly formed open field. However, LOS magnetogram showed that the filament's left footpoint and the nearby open field share the same polarity (Fig. \ref{fig3}(a)). This suggested that this jet is more similar to previously reported nanojets driven by curved magnetic field lines reconnecting at small angles \citep{2021NatAs...5...54A}. This also explains why no post-reconnection closed loops were observed in the $\mathrm{HRI}_{\mathrm{EUV}}$ images, and why this jet was so transient and the filament did not fully reconnect with the open field and evolve into an erupting filament.

\subsection{The failed eruption of the upper filament}
With these jets temporarily stopped, the filament erupted  around 22:00 UT and reached maximum height at about 22:10 UT. Just then, we can clearly observe the existence of a double-layer structure (Fig. \ref{fig1}(d5)). This is a so-called double-decker filament, where two filaments at different heights  share the same polarity
inversion line \citep{2012ApJ...756...59L,2014ApJ...789...93C,2017ApJ...836..160Z,2018NewA...65....7T,2019ApJ...875...71Z,2020ApJ...900...23M,2021ApJ...923..142C,2022ApJ...929...85D,2022ApJ...933..200Z,2023ApJ...959...69H,2024ApJ...961...11K,2024ApJ...961..145Z,2024ApJ...964..125S}. The magnetic field configuration supporting this structure and its possible formation mechanism are analyzed in detail in Appendix \ref{A1}. We analyzed the double-decker filament eruption by placing a slice (the blue dashed line in Fig. \ref{fig1}(d5), parallel to the X-axis and spanning across each layer of the double-decker filament, thoroughly capturing the eruption process) along the more pronounced part of the double-layer structure to obtain an eruption time-space plot. We found that there were two intensity minima along the slice, corresponding to the upper and lower filament (UF and LF). The unprecedented spatial resolution attributed to $\mathrm{HRI}_{\mathrm{EUV}}$ revealed the existence of the UF which is only 2-3 Mm thick. By fitting Gaussians to these two minima, we obtained a distance between the UF and LF of 6.8 Mm (projected height only, Fig. \ref{fig1}(e)).

Interestingly, the UF eruption was unsuccessful; instead, another bidirectional jets flowed along the UF appeared during the UF lifting, resulting in the drainage of UF material. We attributed this to the reconnection of the UF with the fields that wrap around the filament, and the bidirectional jet pushed the filament material towards the footpoints (Fig. \ref{fig1}(d6), especially clear in the movie). In the time-space plot(Fig. \ref{fig1}(f)), we could also find a clear time lag between UF and LF eruption (about 15 min), and the UF projected eruption speed (about 4.9 km $\mathrm{s}^{\mathrm{-1}}$) was significantly smaller than that of LF (about 11.5 km $\mathrm{s}^{\mathrm{-1}}$). This could be due to the UF eruption weakening the overlying field first, or it could be due to the strong kink instability of LF shown during the LF rotational eruption (see the following section). Notably, considering the filament location (W150S18) and the separation angle (83°), the projected speed measurements underestimate the true value by 11\%, assuming a radial eruption trajectory. The whole process is clearly visible in the movie.

\subsection{The failed eruption of the lower filament}
Bidirectional jets led to UF material drainage, causing the UF to become nearly invisible in $\mathrm{HRI}_{\mathrm{EUV}}$ images (Fig. \ref{fig2}(a)). At about 22:15 UT, the LF began to lift. The outer edge of the lifting LF showed a clear brightening and fine features that were indicative of a magnetic flux rope structure. Previous studies suggested that such a brightening was caused by the reconnection of the magnetic flux rope and the overlying field \citep{2024MNRAS.530..473X}. We also observed that the left part of the LF brightened significantly more than the right part (Fig. \ref{fig2}(c)). We argue that this elongated (strip shape) brightening results from the reconnection of the lifting LF with the UF magnetic structure that lost material. The elongated brightening was also consistent with one of the double-decker filament eruption scenarios summarised in previous studies \citep{2024ApJ...961...11K}, namely with an eruption due to a rising lower magnetic flux rope and a merging with higher magnetic flux rope. The LF showed a clearly twisted structure during eruption, further indicating its magnetic flux rope structure. As the LF rotated and lifted further, it interacted with the overlying field (brightening in Fig. \ref{fig2}(d)), and the LF material fell  back along the overlying field. The continuous material fallback also caused brightening at the overlying field footpoints. (see the white ellipses in Fig. \ref{fig2}(e) and movie).

After the failed eruption of LF, a small and compact loop appeared above the LF left part, which we consider to be a typical post-flare loop formed by stretching the overlying field during the LF eruption. Besides this small and compact loop, a large loop gradually became visible above the LF right part (Fig. \ref{fig2}(f)). Previous observational studies have suggested that  the overlying loops were filled by chromospheric material that has been heated and evaporated by the accelerated electrons \citep{2012A&A...548A..89N}. Although we agree with this process, no significant hard X-ray sources were detected by the Spectrometer/Telescope for Imaging X-rays (STIX) \citep{2020A&A...642A..15K} at the loop apex. Meanwhile, we observed distinct falling filament material along the large loop. Therefore, we suggest that the chromospheric material, heated and evaporated by the falling filament material, likely played a dominant role in the large loop's appearance. The formation of this large loop can also be understood as resulting from impact-driven magnetic reconnection, where the falling filament material or contracting magnetic structure triggers reconnection with the low-lying field, similar to the mechanism proposed by \cite{2020ApJ...905..126W} for contracting filaments.

We extracted normalized lightcurves for the small loop and the left footpoint of the large loop by summing over the red and blue dashed box in Fig. \ref{fig2}(g), respectively. We find that the brightness of the small loop peaks at least 70 minutes earlier than for the large loop, which also suggests a different production mechanism.

\begin{figure}
\includegraphics[scale=0.87]{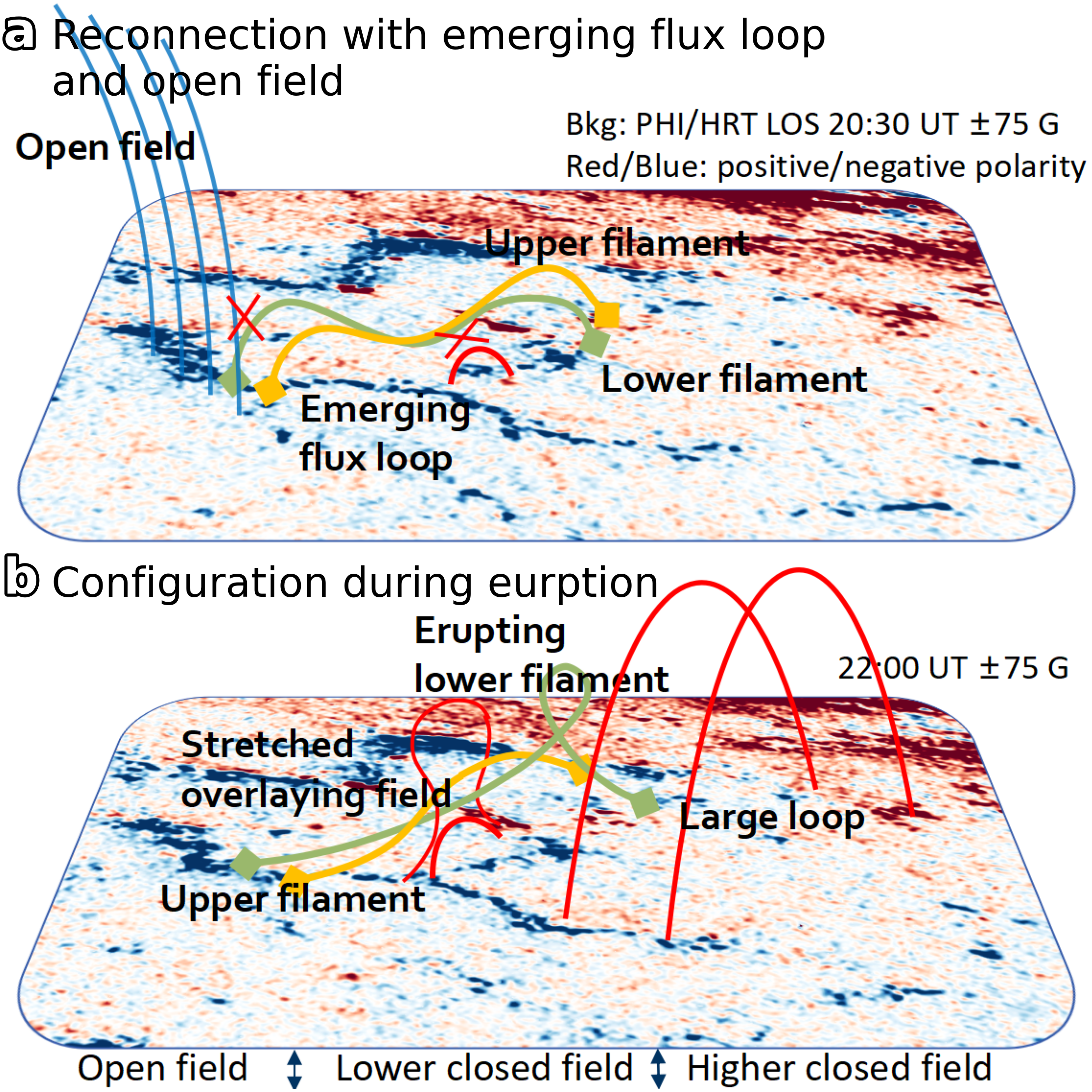}
\caption{\label{fig3}(a): The reconnection process between double-decker filament, nearby open field, and emerging flux loop at the bottom, with orange and green curves for UF and LF, blue curves for nearby open field, respectively. (b): The small post-flare loop formed by the filament lifting stretching the lower closed field and the large loop formed by the falling filament material, with the bottom space divided into three ranges to indicate the magnetic field around the filament.}
\end{figure}

\section{Discussions}
The unprecedented spatial resolution of $\mathrm{HRI}_{\mathrm{EUV}}$ enabled us to study the fine structure of a filament and magnetic reconnection processes during its eruption. Meanwhile, the HRT LOS magnetogram further confirmed the magnetic configuration, which we used as the background for a cartoon illustration (Fig. \ref{fig3}).

The double-decker filament was activated by the emerging flux loop below it, and the left part of the filament was then reconnected with the nearby open field, triggering  bidirectionally propagating collimated jets (Fig. \ref{fig3}(a)). Subsequently, reconnection with the wrap-around field during the UF lifting generated other bidirectional jets, leading to material drainage from the UF (wrap-around field not shown in the cartoon; see Fig. \ref{fig1}(d6) and movie). The more extended right part of the LF showed strong twisting and rotation during eruption and reconnected with the higher overlying field on the right side, leading to the fall of the filament material along the overlying field. At the same time, the left part of the LF (including UF) lifting stretches the lower overlying field on the left to form the small classical post-flare loop (Fig. \ref{fig3}(b)).

Recent studies have suggested that mini-filament (i.e., scale of about 10 Mm) eruptions, which act as a scaled-down version of large-scale (100-1000 Mm) filament eruptions \citep{2011ApJ...735L..43S,2015Natur.523..437S,2020ApJ...896L..18S,2023FrASS..1017870S,2024ApJ...969...48P}, may be the universal mechanism for triggering solar jets and may explain the solar wind sources \citep{2023ApJ...945...28R,2023Sci...381..867C,2024ApJ...963....4S,2024ApJ...965L..18F}. The 
current filament was of medium scale (40 Mm), which means that it potentially reflects the dynamics of both mini and large-scale filament eruptions. Moreover, it has a double-layer structure and was surrounded by a complex magnetic field configuration, covering a wide range of forms from open to closed field, with the left (right) half of the filament reconnecting with the open (closed) field. We therefore observed the simultaneous generation of coronal jets and failed eruptions from a single filament eruption. Further, due to the double-layer structure of the filament and the inhomogeneity of the closed field, we could also observe both the classical post-flare loop and the larger flare loop above the filament in a single eruption.
This suggests that filaments, as common structures in the solar atmosphere, can trigger eruptions on different scales by interacting with the surrounding magnetic field through the universal process of magnetic reconnection. The filament in our study possesses medium-scale, with its eruption encompassing both typical large-scale features and fine structures. This demonstrates the observational feasibility of applying classical magnetic reconnection and MHD instability analysis to mini-scale filaments, which relies on high-resolution observations from new-generation instruments represented by Solar Orbiter.

Beyond triggering filament eruptions, magnetic reconnection is also a major cause of failed or partial eruptions, occurring either through reconnection between the two filament legs (due to rotation caused by strong kink instability during the eruption \citep{2003ApJ...595L.135J,2005ApJ...630L..97T,2012A&A...548A..89N,2019ApJ...877L..28Z}) or through reconnection between the erupting filament and the overlying field \citep{1999ApJ...515L..81A,2012A&A...548A..89N,2015ApJ...808L..24C,2016IAUS..320..221G,2020ApJ...889..106Y,2020ApJ...904...15Y,2022ApJ...933L..38D,2023ApJ...959...67C,2023MNRAS.525.5857J,2024MNRAS.530..473X,2024SoPh..299...81M,2024SoPh..299...89X,2024A&A...687A.130L}. In traditional models, reconnection events are typically large-scale and catastrophic, potentially determining a filament's fate in a single event. 
By analyzing the entire eruption event and comparing it with traditional eruption models, we discovered a previously unseen, dynamic, and nonmonotonic process: the filament continuously undergoes small-scale reconnection events, each modifying its magnetic field and material configuration. We define this process as "persistent magnetic cutting", which inherits from the traditional tether-cutting reconnection concept but emphasizes the continuous cutting process occurring over an extended period before eruption, characterized by four core characteristics:
\begin{itemize}

\item Persistence: Unlike the catastrophic reconnection occurring at specific stages in traditional theory, small-scale reconnection events persist throughout the filament's evolution, resulting in a gradual evolutionary pattern.
\item Scale: Individual reconnection events are much smaller than the catastrophic ones in traditional models.
\item Configuration: Unlike traditional theory focusing on reconnection at specific locations, persistent magnetic cutting involves reconnection at multiple sites, interacting with various magnetic field structures (emergent flux loops, open fields, wrap-around fields, etc.), exhibiting more complex spatial distribution features.
\item Accumulation: The accumulative effect of these frequent small-scale reconnection events significantly impacts filament stability, fundamentally different from the single catastrophic reconnection in traditional theory.
\end{itemize}

Given the complexity of the coronal environment, this persistent process of magnetic reconnection significantly influences eruption success, particularly for small- and medium-scale filaments. This persistent magnetic cutting effect may explain why many small- to medium-scale active region filaments have much shorter lifetimes compared to quiescent ones: the ubiquitous magnetic reconnection in the corona, which is difficult to capture with low-resolution instruments, greatly affects filament stability. Many stable or erupting small- and medium-scale filaments may simply dissipate quietly into the solar atmosphere due to this ubiquitous magnetic reconnection.

\section{Conclusions}

Using unprecedented high-resolution observations from Solar Orbiter, we identify complex interactions between a double-decker filament and its surrounding magnetic environment, exhibiting persistent and cumulative effects that differ from the single catastrophic reconnection events in traditional theory (see Appendix \ref{A2} for detailed comparison). Our proposed "persistent magnetic cutting" provides a new perspective for understanding filament stability: persistent small-scale reconnection gradually alters the filament's magnetic topology and material distribution. Although quantitative analysis of this effect based solely on observations remains challenging, future research combining advanced numerical simulations will enable deeper investigation of this physical process.
This concept bridges the gap between mini-scale filaments (driving coronal jets) and large-scale filament eruptions (driving CMEs), demonstrating the dominant role magnetic reconnection plays in filament evolution across different scales. 

\section{Data availability}
Solar Orbiter data are publicly available through the Solar Orbiter Archive\footnote{https://soar.esac.esa.int/soar/}. This research used the SunPy \citep{sunpy_community2020,Mumford2020} software package to present the observation results. These EUI, PHI and STIX joint observations were part of a Solar Orbiter major flare campaign that included several observing windows in late March and early April 2024.  (See \citep{2020A&A...642A...3Z,2025arXiv250507472R} for an overview of the campaign). The $\mathrm{HRI}_{\mathrm{EUV}}$ operated with long-exposure images (cadence of 16 seconds) paired with short-exposure images (cadence of 2 seconds). We used 897 frames of long-exposure images (19:59-23:59 UT) from the $\mathrm{HRI}_{\mathrm{EUV}}$ level-2 data \citep{euidatarelease6}. The cross-correlation technique (see \citep{2022A&A...667A.166C} for details) was applied to reduce  jitter when generating the time-space plots in Fig. \ref{fig1}(f). Simultaneously, PHI/HRT obtained 8 full polarimetric frames (20:00-23:30 UT) with a 30-minute cadence, providing crucial magnetic field information throughout the eruption. In this work, only the LOS magnetic field was used.

\begin{acknowledgements}

Solar Orbiter is a space mission of international collaboration between ESA and NASA, operated by ESA. The EUI instrument was built by CSL, IAS, MPS, MSSL/UCL, PMOD/WRC, ROB, LCF/IO with funding from the Belgian Federal Science Policy Office (BELPSO); the Centre National d’Etudes Spatiales (CNES); the UK Space Agency (UKSA); the Bundesministerium f\"ur Wirtschaft und Energie (BMWi) through the Deutsches Zentrum f\"ur Luft- und Raumfahrt (DLR); and the Swiss Space Office (SSO). The German contribution to PHI is funded by the BMWi through DLR and by MPG central funds. The Spanish contribution is funded by AEI/MCIN/10.13039/501100011033/ and European Union “NextGenerationEU”/PRTR” (RTI2018-096886-C5,  PID2021-125325OB-C5,  PCI2022-135009-2, PCI2022-135029-2) and ERDF “A way of making Europe”; “Center of Excellence Severo Ochoa” awards to IAA-CSIC (SEV-2017-0709, CEX2021-001131-S); and a Ramón y Cajal fellowship awarded to DOS. The French contribution is funded by CNES. The STIX instrument is an international collaboration between Switzerland, Poland, France, Czech Republic, Germany, Austria, Ireland, and Italy. The AIP team was supported by the German Space Agency (DLR), grant numbers \mbox{50 OT 1904} and \mbox{50 OT 2304}. Y.S. was supported by the  Shenzhen Key Laboratory Launching Project (No. ZDSYS20210702140800001) and the Specialized Research Fund for State Key Laboratory of Solar Activity and Space Weather.
\end{acknowledgements}

\bibliographystyle{aa} 
\bibliography{ref} 

\begin{appendix}

\section{The magnetic configuration and filament double-layer structure origin}\label{A1}

\begin{figure*}
\includegraphics[scale=0.6]{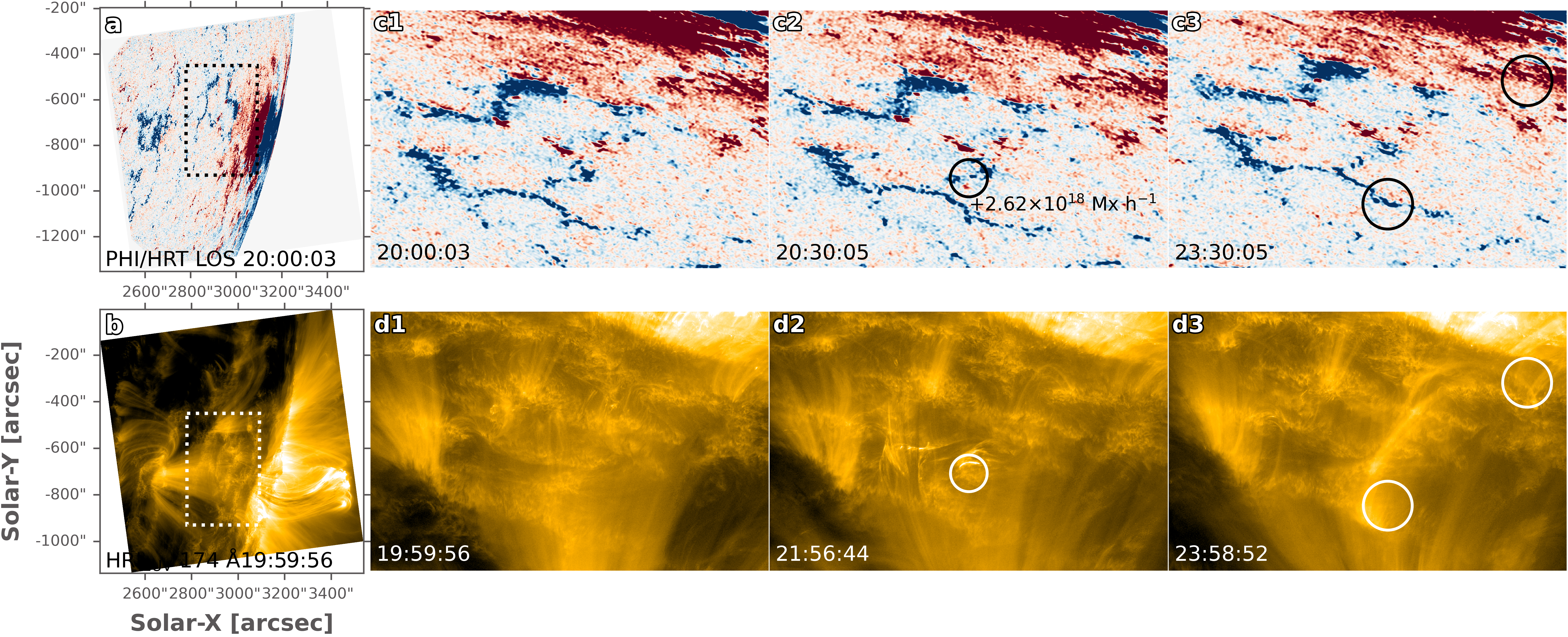}
\caption{\label{fig4}: Comparison of PHI/HRT LOS magnetic field data with HRI observations. The emerging flux loop and its corresponding emerging flux region are marked in panels (c2) and (d2). The footpoints of the large loop are annotated in panels (c3) and (d3). The
 PHI/HRT maps  were shifted (+15$''$, -140$''$) to align with HRI map based on identifying characteristic features.}
\end{figure*}

PHI/HRT temporal resolution for vector data can be as short as 1 minute, but, due to telemetry limitations, a 30-minutes cadence was used for the Major Flare SOOP. Here, we focused on qualitative magnetic topology analysis rather than quantitative measurements. As shown in Fig. \ref{fig4}(a and b), the unrotated HRT and HRI maps shared similar FOVs but exhibited coordinate offsets, requiring manual alignment. Our alignment approach relied on identifying characteristic features. For instance, in panels Fig. \ref{fig4}(c1) and (d1), we found that the filament's left footpoint and the nearby open field shared the same negative polarity. The nearby open field here could also be the bottom of large-scale coronal loops that do not really extend to interplanetary space, but can effectively be considered as open field, given their size relative to the filament. Further confirmation comes from comparing the footpoints of the eventually visible large loop, allowing us to identify its positive and negative footpoints in Fig. \ref{fig4}(c3). Notably, in Fig. \ref{fig4}(c2), we detected weak positive polarity emergence compared to the initial LOS magnetogram, coinciding with the location of the emerging flux loop in Fig. \ref{fig4}(d2). This observation supports our interpretation that flux emergence beneath the filament triggers jets along it. Meanwhile, we calculated the emergence rate of positive flux within the black circles in Fig. \ref{fig4}(c1) and (c2), corresponding to the emerging flux loop region. These findings demonstrate the crucial role of high-resolution magnetic data in studying  filament fine-scale evolution.

Furthermore, with the magnetic field configuration confirmed, let us turn our attention to this filament's interesting double-layer structure. The origin of double-decker filaments is an open question. In previous studies, these  structures have been generally considered  to originate from the splitting of a single filament \citep{2012ApJ...756...59L,2018NewA...65....7T,2022ApJ...929...85D,2022ApJ...933..200Z}. More recently, it has been suggested that they may arise from the successive eruption of two magnetic flux ropes \citep{2024ApJ...964..125S}. Here, we examine this question by analyzing jet motion through the filament structure. There were multiple jets during the filament eruption (Fig. \ref{fig1}(c)), and we found that the initial jet was very slender and concentrated. However, before the filament erupts, the jet had already bifurcated (Fig. \ref{fig1}(d1)), and this remains true after the filament eruption as well. By tracing the bifurcation of jets generated at the filament footpoints, previous study found that the coronal cavity and the filament are different parts of a single magnetic flux rope, with the two components separated in height \citep{2022MNRAS.516L..12T,2006JGRA..11112103G}. Following this idea, we argue that continuously emerging flux loops reconnect with the filament, leading to the splitting of the filament, and that the jets generated by the reconnection exhibit bifurcations along the newly formed double-layer structure, respectively. However, high temporal magnetic field data are currently not available, and we will continue to discuss this question in future studies.

\section{Magnetic reconnection in the heliosphere system}\label{A2}

Finally, let us place the reconnection term in a broader heliospheric context. Magnetic flux rope erosion, a key process affecting CMEs during their interplanetary propagation, was first identified through observations of interaction between flux ropes and ambient solar wind magnetic fields \citep{1987GeoRL..14..355G}. A significant advancement came when it was demonstrated quantitatively how magnetic reconnection removes flux from the rope's outer layers and creates open field lines \citep{2007SoPh..244..115D}. Further characterization of this process provided detailed analysis of reconnection locations and erosion rates. Studies revealed that the majority of erosion (47-67\%) occurs within 0.39 AU of the Sun \citep{2015JGRA..120...43R}, highlighting the importance of inner heliosphere dynamics. More recent work has shown how this erosion process varies with the solar cycle and affects the geoeffectiveness of CMEs \citep{2020GeoRL..4786372P,2021A&A...650A.176P}, adding another layer to our understanding of this complex phenomenon.

The filament persistent magnetic reconnection proposed in this work is distinct from flux rope erosion (previously widely used in space physics) in several key aspects:
\begin{itemize}

\item Scale: Persistent magnetic reconnection involves multiple small-scale reconnection events ($10^{1}-10^{2}$ Mm for the filament-scale in our event), while flux rope erosion typically occurs through large-scale reconnection ($10^{4}-10^{5}$ Mm for typical interplanetary magnetic-clouds-scale).
\item Location: Filament persistent magnetic reconnection takes place in the low corona before and during eruption, whereas flux rope erosion happens during interplanetary propagation.
\item Reconnection characteristics: Persistent magnetic reconnection exhibits frequent, distributed small-scale reconnection events around the structure, compared to the primarily front/rear localized reconnection in flux rope erosion.
\item Effects: Persistent magnetic reconnection influences the structural stability and eruption dynamics of filament,  while flux rope erosion mainly affects magnetic flux content and helicity.

\end{itemize}
These differences suggest that magnetic reconnection processes in the solar atmosphere and the solar wind operate across multiple scales and regimes, with distinct physical mechanisms and consequences for solar eruptions.

\end{appendix}

\end{document}